\documentstyle[11pt,my]{article}

\begin{document}                                                                                   
\title{Two Decades of Hypergiant Research} 
\author{Alex Lobel}  
\affil{Harvard-Smithsonian Center for Astrophysics, 60 Garden Street, Cambridge 02138 MA, USA}
\date{April 17, 2001}

\begin{abstract}
This article is a brief review of the research by Dr. C. de Jager and co-workers over the past twenty 
years into the physics of hypergiant atmospheres. Various important results on the microturbulence, mass-loss,
circumstellar environment, atmospheric velocity fields, and the Yellow Evolutionary Void of these enigmatic stars
are summarized. Aspects of recent developments and future work are also communicated.      
\end{abstract}



\section{Introduction}  
In 1980 Prof. Cornelis de Jager at the University of Utrecht published a book entitled {\em The Brightest Stars}. 
It followed years of scientific research into the question of why there are 
no stars brighter than a certain upper limit. In other words, what 
mechanism(s) determine(s) the upper limit of stellar atmospheric stability?
His monograph (de Jager 1980) is a compilation which addressed nearly every aspect of massive star
research, and which has formed the basis of fundamental progress about these questions 
during the past two decades. This `festschrift' reviews some of the (partial) 
answers which have been proposed in the course of time by Dr. de Jager and many co-workers, 
however without attempting to be complete in all the scientific or chronological details. We 
also highlight the most important recent developments, based on a limited number of selected references 
to their work, which will enthuse the reader with this remarkable subject in astrophysics. 

Cool hypergiants ($\rm Ia^{+}$ or $\rm Ia0$) are a class of supergiant stars with $T_{\rm eff}$  below $\sim$10,000 K.
They are physically different from supergiants of luminosity class $\rm Ia$. The usual $\rm Ia^{+}$ 
designation does not mean that they are always more luminous than the former, but they 
differ in their spectral properties. These stars show one or more broad emission components 
in the Balmer H$\alpha$ line profile, which is a signature of an extended atmosphere or 
of a relatively large rate of mass loss. Another important spectroscopic aspect 
of hypergiants is that their absorption lines are significantly broader than those 
of the $\rm Ia$ stars of similar spectral type and luminosity (de Jager 1998). $\rho$~Cas and HR~8752 
are the prototype objects of the yellow hypergiants (F$-$G $\rm Ia^{+}$). They are among the most 
massive (20$-$40~$M_{\odot}$) cool stars presently known. These stars exist near the Eddington 
luminosity limit, and exhibit a wide range of uncommon stellar properties. Their atmospheres 
are unstable, which causes quasi-periodic pulsation variability, strongly developed large-scale 
velocity fields, excessive mass-loss, and extended circumstellar envelopes. 

We summarize three fundamental questions for the study of hypergiants: \\
$i$. Why do hypergiants of similar bolometric luminosity and spectral type as normal supergiants
display the signatures of strong large-scale photospheric motions and enhanced mass loss? Why  
do these properties appear to be absent for normal supergiants? \\
$ii$. What is the evolutionary status of the yellow hypergiants, and how do they fit into 
the broader scheme of massive star evolution? Are they the likely progenitors of 
(Type II) supernovae in the not too distant future? \\
$iii$. Can an understanding of their atmospheric dynamics and circumstellar environments 
provide clues about the origin of the luminosity boundary for evolved stars? Do these 
dynamics characterize the hypergiants uniquely, or do they occur modified in other classes of 
luminous stars as well? \\ 

\section{Microturbulence}

A few years after {\em The Brightest Stars} de Jager (1984) published a theoretical paper
which marked the beginning of a thorough study about the stability of hypergiant atmospheres,
and which has consequently been pursued by many other workers in this field. 
It is put forward for the first time in this paper that atmospheric dynamics is a basic 
ingredient for understanding a temperature-dependent maximum luminosity limit extending across 
the entire upper Hertzsprung-Russell (HR) diagram. The classical Eddington limit is defined 
as the line in the HR-diagram where gravity equals the force produced 
by the radiation pressure due to electron scattering. 
Neglecting other atmospheric opacity sources, this results in a nearly horizontal line in the HR-diagram for stars with 
a single mass-luminosity relation. Hence, the classic limit cannot explain the absence of cool supergiants 
with $L_{\star}$$\gtrsim$$10^{6}$~$L_{\odot}$, because hot stars that are more luminous do exist.
De Jager proposed instead that the local stability of these extended atmospheres is greatly altered by an outward 
acceleration due to the gradient of turbulent pressure which results from the local 
field of turbulent motions. The importance of the latter
scales with luminosity. It causes the effective acceleration, which is due to the balance 
between gravity, radiation- {\em and} turbulent-pressure, to produce higher mass-loss rates 
for the intrinsically brighter stars. The influence of turbulence is particularly important for 
the cool supergiants for which the radiation pressure remains rather small. For this reason
the theoretical luminosity limit of cool stars is occasionally termed the `de Jager limit'
(for turbulent pressure). 

Studies of strong `microturbulence' observed in the spectra of supergiants continue as
an important subject of hypergiant research for obvious reasons. It increases the equivalent width 
of Fraunhofer lines by stochastic small-scale motions with wavelengths smaller than 
the optical scale height of the atmosphere. The line-of-sight velocity component of these movements
samples only a limited part of the photosphere over which the spectral lines are formed. 
A pertinent problem in pinpointing the physical nature of stellar microturbulence is 
that it is often observed to exceed the sound speed in supergiant atmospheres. 
Microturbulence generally increases towards the hotter and more luminous 
stars. It is frequently larger than the isothermal velocity of sound in the Ia and Iab supergiants.

In 1991 de Jager chaired an international congress in Amsterdam about instabilities in
evolved super- and hypergiants. At the conference he presented the idea that 
stellar microturbulence is strongly related to the physics of atmospheric shock waves.  
Through numerical integrations of the hydrodynamics equations an observable microturbulent 
velocity is computed, assuming that the dependence of shock wave energy on the wavenumber is of 
the Kolmogoroff-type (a power law spectrum) (de Jager 1992a). Shock trains propagating outwards are a natural consequence of 
acoustic energy generation in cool stars with a subphotospheric convection zone. De Jager et al. (1997) 
subsequently showed that the thin heated layers trailing these shocks constitute the main part of 
the observed micro-`turbulence'. 
A complex of successive hot sheets, distributed stochastically over the atmosphere, yields a
stochastic distribution of thermal motions that, observationally, cannot be distinguished 
from a random field of small-scale hydrodynamic motions (de Jager 1998).

The application of this theory to the yellow hypergiant $\rho$~Cas revealed that a field of 
atmospheric shock waves which is critically supersonic, or have a mean Mach number only slightly exceeding unity,
reproduces the observed supersonic microturbulence velocity of 11~$\rm km\,s^{-1}$. 
Remarkably, the hydrodynamic velocity field in the shocks remains very small, of the order of 
only 0.5~$\rm km\,s^{-1}$. Nevertheless, the momentum provided by such weak shock wave trains, assuming 
spherical geometric symmetry, is sufficient to initiate the large mass-loss rates (typically 
a few times $10^{-5}$~$M_{\odot}$$\rm yr^{-1}$) observed in cool hypergiants (de Jager 1996). 
These rates partly result from the very large stellar surface area  (the mean photospheric radius of $\rho$~Cas ranges between 
400 and 500~$R_{\odot}$), and the sonic point which is situated well inside the photosphere (de Jager et al. 2001). 
The field of stochastic shock waves provides on average outward directed momentum which diminishes the effective atmospheric 
acceleration and increases the atmospheric scale height.   

\section{Mass Loss}

Evolutionary calculations show that yellow supergiants evolve along tracks of nearly constant luminosity, 
beneath an observational luminosity limit. This limit has been called the Humphreys-Davidson (HD) limit,
when it was pointed out that the luminosity of the brightest blue stars is about a factor six 
higher than that of the luminous red stars. Very evolved luminous supergiants move {\em bluewards} 
in the HR-diagram because they shed copious amounts of mass to the circumstellar/interstellar environment 
in the red supergiant phase. During redward evolution the very high mass-loss rates ($\dot{M}$) reduce the stability 
of the convective layer, and below a critical envelope mass it contracts into a thinner radiative envelope, 
which causes rapid blueward evolution. For only a limited range of initial masses stars can become 
red supergiants, and evolve back far bluewards. The late stages in the fast evolution of these 
massive stars are relatively short. Typical values computed for stars with masses between 20 and 40~$M_{\odot}$ 
are around one million years, with blueward tracks which last only a few times $10^{4}$ years. 
The high $\dot{M}$ for very massive stars is however thought to have a decisive influence on these late evolutionary stages.     

De Jager et al. (1988) presented a comprehensive table of empirical mass-loss 
rates for 271 objects all over the HR-diagram, together with a parameterization  
which holds for stars from early to late spectral types (Nieuwenhuijzen \& de Jager 1990).
$\dot{M}$ primarily depends on the stellar luminosity $L_{\star}$. For hot stars there appears 
to be a rather linear dependence, whereas K and M stars deviate by about 3 dex from the average 
hot star dependence. This points to other parameters than $L_{\star}$ (and $T_{\rm eff}$)
which influence the mass-loss rate (de Jager 1993). It seems imperative to ascribe the difference 
to the different physical mechanisms which drive the massive stellar winds in the hot and 
in the cool supergiants. These differences are apparent from the emission line 
components observed with high dispersion in $\rho$ Cas at either or both sides of the Doppler displaced
H$\alpha$ absorption core (Lobel et al. 1998). These complex line profiles indicate (quasi-) cyclic variations of the atmospheric 
circumstances and dynamics, which causes variable mass-loss rates that can occasionally increase to extreme 
values of a few $10^{-4}$~$M_{\odot}$ $\rm yr^{-1}$. The temporal variability observed in 
photospheric absorption lines is not correlated with the changes of H$\alpha$; the latter is 
at least twice longer than the former. Since H$\alpha$ forms higher, over a larger fraction 
of the atmosphere, it indicates that the atmosphere is dynamically velocity stratified, and the pulsation 
strongly enhances the density scale height compared to hydrostatic models.  

Yellow hypergiants exhibit `eruptions', which occur on time-scales much longer than 
the pulsation quasi-period (ca. half a century, say). In an outburst of 1945-46 $\rho$~Cas suddenly 
dimmed and displayed TiO-bands in its spectrum, characteristic of the photospheric temperatures of M-type stars. 
Within a couple of years the star brightened up by nearly a magnitude (April '47), and a mid G-type spectrum was 
recovered around 1950. In 1985-86 the star showed a larger-than-average amplitude in the light curve, which can
be associated with shell ejection events. Pulsation variability seems to be a common 
phenomenon in these stars. In his review on yellow hypergiants de Jager (1998) hypothesized 
that a distinction should be made between the periods in which these stars can be considered as fairly 
quietly pulsating, and when they are in the `explosive' phase. During that phase the 
effective atmospheric acceleration practically vanishes, resulting in strongly enhanced
mass ejection. Interestingly, it is suggested that the mechanism for outburst may be related 
to increased pulsation velocities during a period of instability that occurs when the 
star, during rapid blueward evolution, increases its pulsation amplitudes. The broader
view emerges here that the exceptional atmospheric dynamics manifested by the yellow hypergiants
is intricately related to their late evolutionary stage. 

\section{Circumstellar Environment}

Little is known about the circumstellar environments of yellow hypergiants.
The mid-IR silicate emission feature of $\rho$ Cas is very weak, indicating the formation of dust only at very large 
distance from the surface. However, IRC$+$10420 is of later spectral type and possesses a very extended dust shell. 
It is commonly assumed that mass-loss from cool massive stars is primarily spherically symmetric.
The absence of observational evidence for strong magnetic fields in these supergiants and slow 
rotation does not readily support wind flows into a preferential direction. It is know 
that for most if not all stars wind flow is not spherically symmetric. Magnetic fields and 
rotation have a strong tendency to concentrate the gas that is leaving the star towards the 
equatorial plane, thus producing a disk rather than a spherical wind structure (de Jager 1993).

HST-WFPC images of IRC$+$10420 reveal a circumstellar environment with a bipolar symmetry, 
and jet-like features and arcs produced primarily in the equatorial region. A bipolar wind has also been 
proposed for $\rho$ Cas (Lobel 1997), based on the phenomenon of line core splitting 
observed in optical absorption lines of very low excitation energy. $\rho$~Cas' emission line spectrum is excited 
in the circumstellar environment with an excitation temperature of 3050~K.
This emission spectrum can emerge from a fast bipolar wind which collides with denser circumstellar 
(or perhaps interstellar) material, expelled during its violent mass-loss history. Similar split line cores 
were also observed for HR~8752 in 1960-70, when it was the spectroscopic twin of $\rho$~Cas.   
The line core splitting phenomenon is of a bewildering complexity, because the central reversals
are permitted emission lines which appear in the core of metal absorption lines of resonance {\em and} subordinate 
transitions. $\rho$ Cas' emission spectrum is not excited by a normal stellar chromosphere because 
its spectrum lacks the classic chromospheric indicators, except for H$\alpha$. Long-term spectroscopic monitoring 
of the split profiles reveals that the central emission peak is static, while the intensity of either of the 
adjacent absorption components periodically intensifies and weakens. This supports the formation 
of the emission line spectrum at large distance from the star, where it does not share the 
photospheric pulsation movements below. The blue-shifted appearance of the central emission cores 
indicates their formation in an optically thick wind. On the other hand, we have also discovered forbidden 
[Ca~{\sc ii}] emission lines in the spectrum, which reveals a very extended tenuous 
gas envelope surrounding $\rho$ Cas. Prominent [N~{\sc ii}] line emission is also observed in HR~8752.                   

Surprizingly however, the same metal core splitting phenomenon is observed in FU Ori objects. 
These young stellar objects have bipolar outflows, produced by disk accretion, with powerful outbursts 
of mass-loss. They are pre-main sequence objects with optical spectra resembling F- and G-supergiants.
Interestingly, it has also been proposed that FUor eruptions are due to perturbations in a circumstellar 
disk caused by the passage of a compact companion on an eccentric orbit. There is presently 
no indication for a compact companion in $\rho$ Cas, but clear evidence for an early B-type main sequence 
star in HR~8752 was obtained from continuous near-UV spectra observed with $IUE$.   

Whether split absorption profiles are the signature of circumstellar disks which can obstruct strong wind outflow 
remains currently unanswered and requires further research. It appears however that they are spectral indicators 
of complex and dynamic circumstellar environments, which characterize the yellow hypergiants.  

\section{Atmospheric Velocity Fields}

In 1990 de Jager published a short paper in {\em Solar Physics}. The paper is remarkable 
in that it provides an explanation for the problem of the `granulation boundary' in the HR-diagram. 
This boundary has been inferred from the shape of spectral absorption line bisectors which incline 
oppositely when moving from the lower to the higher effective temperature regions in the HR-diagram.
For the Ib supergiants the bisector turnover occurs around spectral types G0-5, but is poorly defined 
for the more luminous Ia and $\rm Ia^{+}$ stars. The spectra of F-type supergiants display 
uncommonly broad absorption lines which are ascribed to a well developed large-scale velocity field.
The large macrobroadening cannot be attributed to rapid rotation (i.e. large $v$sin$i$-values) for 
these evolved and very extended stars. For example, the modeling of line profiles in $\rho$~Cas yields 
highly supersonic macroturbulent velocities of $\sim$25~$\rm km\,s^{-1}$. De Jager (1990) proposed 
that for stars to the left of the boundary, in the optically thin regions of the atmosphere 
above the convection zone, internal gravity waves develop which involve a large part of the line forming region. 
To the right of the granulation boundary the atmospheres are in convective equilibrium, and gravity
waves would only affect the smallest optical depths, which hardly influences the line profiles. 

In a major paper which was published the year after, de Jager et al. (1991)
provided the mathematical and physical foundation for these insights. The most 
important wavelengths in the spectrum of gravity waves are close to the atmospheric 
cut-off wavelength. The latter results from radiative damping and the atmospheric 
curvature. It is pointed out that in the highest atmospheric levels only the 
very low-mode (long wavelength) gravity waves are possible. For supergiant pulsation, 
high $l$ modes are not expected because the wavelength of internal gravity waves 
propagating in these extended atmospheres should exceed this cut-off wavelength.
In deeper layers the damping cut-off wavelength becomes gradually shorter and higher 
gravity waves become possible. The more extended the atmosphere the lower the 
mode-number of gravity waves that may develop.  However, gravity 
waves cannot propagate inside convectively stable regions where their 
minimum period (the Brunt-V\"{a}is\"{a}l\"{a} period) becomes real and non-zero.
Gravity waves propagate mainly in the horizontal direction, along the stellar circumference.
They do not propagate strictly vertically. 

The application of this theory (de Jager 1998) to the yellow hypergiants $\rho$ Cas and HR~8752, 
and to the white hypergiant HD~337579, provided a number of impressive theoretical results.
Based on the {\em diagnostic diagram}, which plots the periods of pressure- and gravity-waves 
against their respective spectra of possible wavelengths and the observed quasi-period(s), 
he showed that these atmospheres are dominated by systems of gravity waves with in some cases 
shock waves superimposed. For $\rho$ Cas the long-period waves are due to gravity 
waves of the shortest allowed wavelengths and periods, with fairly high $l$-values. 
The short-period components show the presence of pressure waves with wavelengths 
of about half the stellar radius. $l$-values exceeding unity are consistent with the 
finding for $\rho$~Cas (Lobel et al. 1994), from an improved Baade-Wesselink pulsation test,
that photospheric {\em radial} pulsations must be discarded to combine $T_{\rm eff}$-changes 
with the co-eval radial velocity- and light-curve, observed over a complete variability period.
The quasi-periodicity observed in the light-curves of cool hypergiants therefore results from 
a complex, but law-abiding, interplay of atmospheric (non-radial) pulsation modes due to gravity 
and acoustic waves.  

\section{The Yellow Evolutionary Void}

In 1995 Nieuwenhuijzen \& de Jager published a classic paper on the atmospheric 
accelerations and stability of supergiant atmospheres. Their research provides
a well-founded theoretical explanation for the conspicuously sparse amount of white (late A to early F-type) 
supergiants in the HR-diagram. At the Amsterdam conference de Jager (1992b) raised the question of why most of the
known yellow hypergiants appear to cluster around 6500~K $\leq$ $T_{\rm eff}$ $\leq$ 7500~K with different luminosities,
while the area around 10,000~K and log($L_{\star}/L_{\odot}$)$\geq$5.7 is nearly devoid of stars.
Due to the natural deficiency of cool supergiants it is possibly hard to judge the reality of their absence  
around the upturn of the HD limit, but current population studies support only a minimal number of  
very luminous white supergiants. This area was tentatively baptized the `evolutionary void' after detailed 
calculations revealed that hydrostatically stable solutions cannot be computed for the atmospheres 
of {\em blueward} evolving yellow hypergiants at a lower temperature boundary of $T_{\rm eff}$$\simeq$8300~K. 
These calculations involve a time-independent solution of the momentum equation, which considers the 
Newtonian gravity acceleration derived from the evolutionary mass, the gas, radiation and turbulent 
pressure gradients, and the momentum of the stellar wind. The latter requires the observed mass-loss 
rate. An `effective acceleration' for the atmosphere is obtained iteratively, but which becomes 
negative at the cool border of the void. The outwards directed net force causes an unstable 
atmosphere above 8300~K. A remarkable byproduct of this study is that as stars move 
along their track, for time scales longer than the dynamic time scale of the atmosphere,
the atmosphere continuously adapts to the new ($L_{\star}$, $T_{\rm eff}$)-values, and 
remains stable without reaching the Eddington limit. Current practice of determining 
the stability limit by extrapolating hydrostatic models to a `modified' Eddington limit
(i.e. utilizing a flux-mean opacity for the atmosphere which includes line opacity) 
is therefore not justified. When a real star slowly changes its effective gravity 
acceleration to zero during evolution the atmosphere puffs up, thereby decreasing the opacity 
to the classic electron scattering value.  
 
Over the past few years Drs. de Jager and Nieuwenhuijzen commenced a profound investigation 
of several individual cool hypergiants to which their theoretical developments are being applied
(Nieuwenhuijzen \& de Jager 2000). Atmospheric model parameters of HD~33579
have been determined via the `spectral diagnostics method' and reveal that it is a redward 
evolving (high-mass) supergiant inside the void. For such a star the void is not forbidden. 
HR~8752 and IRC$+$10420 have lower masses and evolve bluewards, currently approaching the void. 
Several independent studies have demonstrated a significant overabundance of sodium in $\rho$ Cas.  
It is brought to the photosphere by dredging-up, which would support its post-red supergiant phase.
De Jager \& Nieuwenhuijzen (1997) suggested that the recurrent eruptions in
yellow hypergiants occur when these stars approach the cool boundary of the void, and `bounce off' 
redward (for a graph see Fig. 1 of Nieuwenhuijzen \& de Jager 2000). This bouncing against the 
void may explain why most of the cool luminous hypergiants cluster 
near its low-temperature boundary, while the identification of hypergiants of later spectral type, 
possibly like VY~CMa (of M-type), is seldom. An increase of the photospheric temperature 
of HR~8752 by 3000 K$-$4000 K has been observed based on high-resolution spectra collected over the past 
thirty years. The effective temperature of IRC$+$10420 has been as low as 5270 K and increased 
to $\sim$7000~K in 1992. The natural question arises whether (apparent) evolutionary changes 
on human time-scales result from an actual active reconstruction of the stellar interior, or  
by some hitherto unknown envelope mechanism? Is there a global instability mechanism in these stars
other than stellar pulsation, which changes their spectral types, perhaps triggered by some internal or external 
cause? Is the void an eternal obstacle for their evolution and will the yellow hypergiants eventually fade as 
supernovae, or will they rapidly evolve across the hypergiant void into early A-type supergiants (e.g. the least 
luminous S Dor stars), after having lost large amounts of mass? Time will tell. 

\section{Future Work}

As is the case with all research at the forefront of science, the present-day developments are the 
most exciting. To avoid disclosing research results prior to the main publication (de Jager et al. 2001), 
we summarize some important general aspects of hypergiant atmospheric stability, and the lines along which 
further investigations are planned. New studies show that different areas in the upper HR-diagram can be 
delineated where the effective acceleration in the upper part of the photosphere is negative, or directed outwards. 
Stars like $\rho$ Cas, HR~8752, and IRC$+$10420 are situated close to this line. Another aspect 
of their dynamic stability is that the sonic point (the point where the wind velocity equals the 
sound speed) of the stellar wind is situated inside the photosphere. Matter which is driven beyond
this point escapes freely into space, causing high mass-loss rates. 

Parallel to these developments is the study of the behavior of the adiabatic indices in the 
atmospheres of luminous supergiants. An early study (Lobel et al. 1992) showed that $\Gamma_{1}$
assumes values below 4/3 in low-gravity hydrostatic model atmospheres. This corresponds to zones of 
increased atmospheric destabilization due to partial hydrogen and helium ionization. The study 
is presently further improved by evaluating Ledoux' stability integral $<$$\Gamma_{1}$$>$
for realistic envelope models (Lobel 2001), including the effects of photo-ionization and the dilution of 
radiation pressure in the extended atmospheres of these stars. Important dependencies on the 
atmospheric acceleration and effective temperature have been demonstrated so far.  

Future work will focus on the interrelations between these different stability criteria,
and their various specific implications. It appears to us that considerable progress will 
be accomplished by integrating the fully time-dependent hydrodynamic equations, and incorporating the various 
acceleration mechanisms in the equation of motion, together with the increased atmospheric 
compressibility due to partial LTE- and NLTE-ionization. These simulations could for example model the 
velocity stratification observed for $\rho$~Cas' atmosphere. The further refinement of regions 
of dynamic instability in the upper portion of the HR-diagram, and a continuous spectral monitoring program 
of HR~8752 (de Jager et al. 2001), for which an outburst seems imminent, is in progress. 

To conclude this overview, it is safe to say that our understanding of hypergiants has considerably 
changed and improved over the past two decades. At the same time, lots of new and pertinent questions 
have been raised about these exceptional objects. We owe much of this progress to the dedicated and continuing 
work of one person; Cornelis de Jager, who is a guide and has been the mentor of many researchers in astronomy 
and astrophysics.                              

\begin{figure}
\plotfiddle{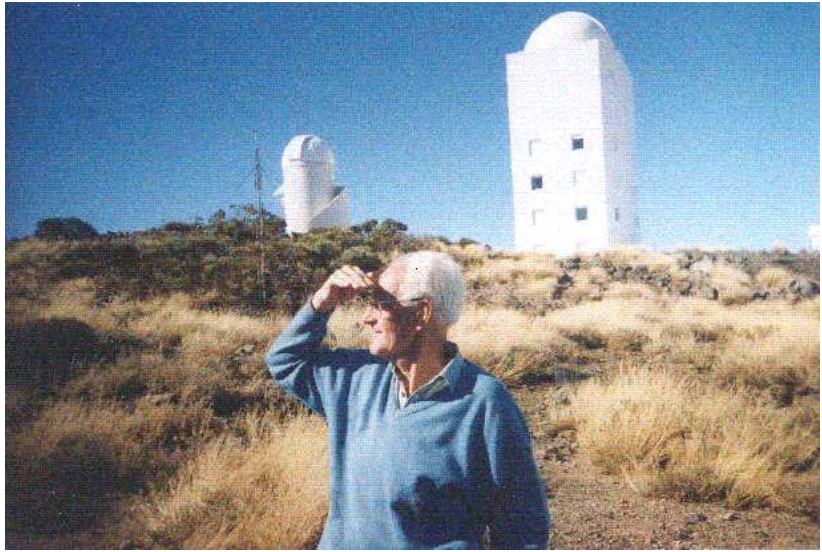}{5.5cm}{0}{50}{50}{-110}{-160}
\vspace*{3cm}
\hspace*{3cm}
\caption{\small Kees de Jager inspecting the Observatori del Teide at Izana on the island of Tenerife (Canary 
Islands, Oct. 1999). The Gregory Coud\'{e} solar Telescope towers in the background.}
\end{figure}

\acknowledgements
This article honors the eightieth birthday of Prof. C. de Jager.  
The author has been a Ph.D. student at the Space Research Organization of the Netherlands - Utrecht 
and the Free University Brussels.

\end{document}